\documentclass[12pt]{article}

\usepackage{scicite}

\usepackage{times}

\usepackage{physics}
\usepackage{graphicx}
\usepackage[labelfont=bf]{caption}
\newcommand{\ks}{\mathbf{k}_{\mathrm{s}}}
\newcommand{\ksa}{\mathbf{k}_{\mathrm{s1}}}
\newcommand{\ksb}{\mathbf{k}_{\mathrm{s2}}}
\newcommand{\ksda}{\mathbf{k}_{\mathrm{s1}}^{\mathrm{D}}}
\newcommand{\ksdb}{\mathbf{k}_{\mathrm{s2}}^{\mathrm{D}}}
\newcommand{\ki}{\mathbf{k}_{\mathrm{i}}}
\newcommand{\kia}{\mathbf{k}_{\mathrm{i1}}}
\newcommand{\kib}{\mathbf{k}_{\mathrm{i2}}}
\newcommand{\kida}{\mathbf{k}_{\mathrm{i1}}^{\mathrm{D}}}
\newcommand{\kidb}{\mathbf{k}_{\mathrm{i2}}^{\mathrm{D}}}
\newcommand{\adagger}{\hat{{a}}^{\dagger}}
\newcommand{\dr}{\mathrm{d}}

\newenvironment{sciabstract}{%
\begin{quote} \bf}
{\end{quote}}

\title{General simulation method \\for quantum-sensing systems}

\author
{Felix Riexinger$^{1,2\ast}$, Mirco Kutas$^{1,2}$, Bj\"orn Haase$^{1,2}$,\\Michael Bortz$^{1}$, and Georg von Freymann$^{1,2}$\\
\\
\normalsize{$^{1}$Fraunhofer Institute for Industrial Mathematics ITWM}\\
\normalsize{Fraunhofer-Platz 1, Kaiserslautern, 67663, Germany}\\
\normalsize{$^{2}$Department of Physics and Research Center OPTIMAS, Technische Universität Kaiserslautern}\\
\normalsize{Kaiserslautern, 67663, Germany}\\
\\
\normalsize{$^\ast$To whom correspondence should be addressed; E-mail: felix.riexinger@itwm.fraunhofer.de.}
}

\date{\today}

\begin{document} 


%\baselineskip24pt


\maketitle


\begin{sciabstract}
Quantum sensing encompasses highly promising techniques with diverse applications including noise-reduced imaging, super-resolution microscopy as well as imaging and spectroscopy in challenging spectral ranges. These detection schemes use biphoton correlations to surpass classical limits or transfer information to different spectral ranges. Theoretical analysis is mostly confined to idealized conditions. Therefore, theoretical predictions and experimental results for the performance of quantum-sensing systems often diverge. Here we present a general simulation method that includes experimental imperfections to bridge the gap between theory and experiment. We develop a theoretical approach and demonstrate the capabilities with the simulation of aligned and misaligned quantum-imaging experiments. The results recreate the characteristics of experimental data. We further use the simulation results to improve the obtained images in post-processing. As simulation method for general quantum-sensing systems, this work provides a first step towards powerful simulation tools for interactively exploring the design space and optimizing the experiment's characteristics.
\end{sciabstract}

% In setting up this template for *Science* papers, we've used both
% the \section* command and the \paragraph* command for topical
% divisions. Which you use will of course depend on the type of paper
% you're writing. Review Articles tend to have displayed headings, for
% which \section* is more appropriate; Research Articles, when they have
% formal topical divisions at all, tend to signal them with bold text
% that runs into the paragraph, for which \paragraph* is the right
% choice. Either way, use the asterisk (*) modifier, as shown, to

\section*{Introduction}
Quantum sensing has become a rapidly evolving field with developments fueled by the outlook of applications that surpass classical limitations \cite{Gilaberte2019}. These range from noise-reduced imaging \cite{Brida2010,Samantaray2017,Sabines-Chesterking:19} over super-resolution microscopy \cite{Classen:17,Unternahrer:18,Tenne2019} to imaging and spectroscopy with undetected photons \cite{Lemos2014,Gilaberte2021,Fuenzalida2022,Lindner21,Kutas2021}. 
Especially one novel quantum-imaging technique -- quantum imaging with undetected photons (QIUP) -- which has been recently experimentally demonstrated provides many potential applications \cite{Lemos2014,Gilaberte2021,Fuenzalida2022}. Using QIUP, an image of an object is obtained without detecting the light that interacted with it. This concept has been adapted to applications in spectroscopy \cite{Kalashnikov2016,Kutas2021} and optical coherence tomography \cite{Paterova2018T,Vanselow2020,Kutas2020}. The principle of these measuring techniques is based on the quantum superposition of the biphoton states of two spatially separated sources \cite{Wang1991,Zou1991}. The light that interacts with the object is never detected. Meanwhile the light that does not interact with the object is used to form an image of that object. The registered image depends on the state of the undetected light due to the superposition of the biphoton states. This distinctive feature of QIUP allows for a probing wavelength at which detection is difficult, while detecting the image at a convenient wavelength \cite{Kutas2022}. 

Some properties of QIUP have been analyzed analytically and numerically \cite{Lahiri2015,Lahiri2019,Gilaberte2021,Fuenzalida2022}. However, each work investigated only isolated aspects of QIUP, while a holistic description considering the interplay of various effects is still missing to date. A detailed understanding is of great importance for the assessment of limiting factors and the potential of this technique. A full model of the system further allows for the exploration of trade-offs between different imaging characteristics such as resolution and visibility. The availability of detailed simulations is becoming more relevant since applications are on the verge of a breakthrough, while their characteristic properties still need to be evaluated experimentally. Using simulations, the design process can be sped up as the system's properties can be explored and optimized virtually.

In this article, we present and demonstrate a method for simulating the detected images of a QIUP setup. This method includes a full model of the biphoton sources and the optical system. With this, a large range of experimental setups including their misaligned states can be simulated.
We further present an approximative variant of the method which speeds up the computation times substantially. For this variant, the model of biphoton sources is simplified. The approximative variant is still applicable to a large range of experimental setups but is limited to fewer types of misalignments. 
We demonstrate our method by simulating detector images and imaging properties of a real experimental setup.
With this, we present a simulation suite enabling the virtual analysis of quantum-sensing experiments for the first time.

\begin{figure}[ht]
    \centering
    \includegraphics[width=\textwidth]{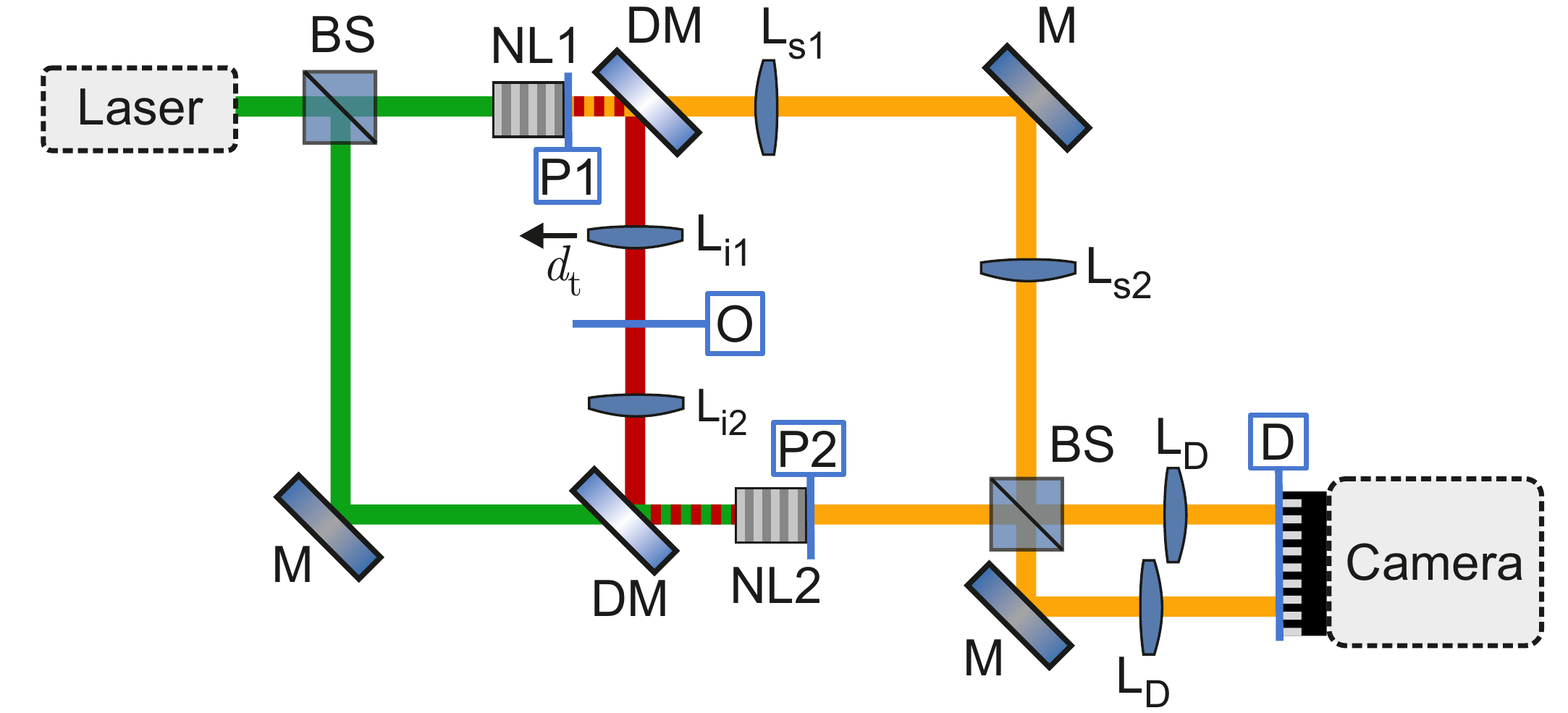}
    \caption{\label{fig:setup}\textbf{Layout of the experimental setup.} The dichroic mirrors separate the wavelengths such that the detected light never interacts with the object. Only the components included in the model are displayed. NL: nonlinear crystal. L: lens. (D)M: (dichroic) mirror. BS: beam splitter. $d_{\mathrm{t}}$ denotes a transversal displacement of the lens. The blue lines denote the source, object and detector planes. Length of optical paths are not to scale.}
\end{figure} 

QIUP schemes with a Mach-Zehnder geometry (Fig.~1) are driven by two coherently pumped identical nonlinear crystals (NL1 and NL2) which produce biphoton states $\ket{\psi_{\mathrm{NL1/NL2}}}$ consisting of a signal (s) and an idler (i) photon. Schemes with a Michelson geometry use one nonlinear crystal through which the pump laser passes twice. We treat the first and second pass as two sources and denote them with NL1 and NL2 to keep the notation consistent. We assume the power of the laser to be small enough, such that stimulated emission can be neglected \cite{Lahiri2019}. The signal parts of the biphoton states emitted from each source (yellow paths) are superposed and subsequently detected with a camera. The idler part (red paths) emitted by NL1 is propagated through NL2 and aligned with the corresponding beam emitted by NL2. In an ideal setup the probability amplitudes and modes emitted by NL1 and NL2 are identical. Therefore, the idler states are indistinguishable and no information about the emitting source can be obtained in the detector plane. Due to the entanglement of the photons s and i, the observed interference on the camera depends on the idler path even though only signal photons are detected \cite{Zou1991,Wang1991,Lemos2014}. To measure an object's properties, it is placed in the path of the undetected idler beam between the two crystals. The transmission coefficient and the phase introduced by the object can then be reconstructed from the detected image \cite{Lahiri2015}.

\section*{Results}\label{sec:results}

\subsection*{Theory of quantum imaging}
Our simulation model consists of three parts: the source model, the propagation model and the detection model. The source model describes the quantum states created by the biphoton sources. The propagation model calculates the propagation of these states through the optical system. The detection part evaluates the interference of the two paths and the resulting signal in the detector plane. In the following, we will limit ourselves to the description of QIUP setups in a far-field configuration.

The source model is used for the calculation of the transition amplitude $g(\ks, \ki)$ for each pair of photon momenta $\ks$ (signal) and $\ki$ (idler). The quantum state of light generated by the sources NL1 and NL2 is given as
\begin{equation}
     \ket{\psi_{\mathrm{NL1/NL2}}} = \int \mathrm{d}\ks \mathrm{d}\ki g_{1/2} (\ks, \ki) \ket{\ks,\ki}
\end{equation}
where $g_{1/2} (\ks, \ki)$ are the transition amplitudes for the nonlinear crystals NL1 and NL2, respectively. We omit the polarization component of the biphoton state for simplicity. Many biphoton sources used for QIUP emit mainly one combination of photon polarizations.

\begin{figure}[htb]
    \centering
    \includegraphics[width=0.9\textwidth]{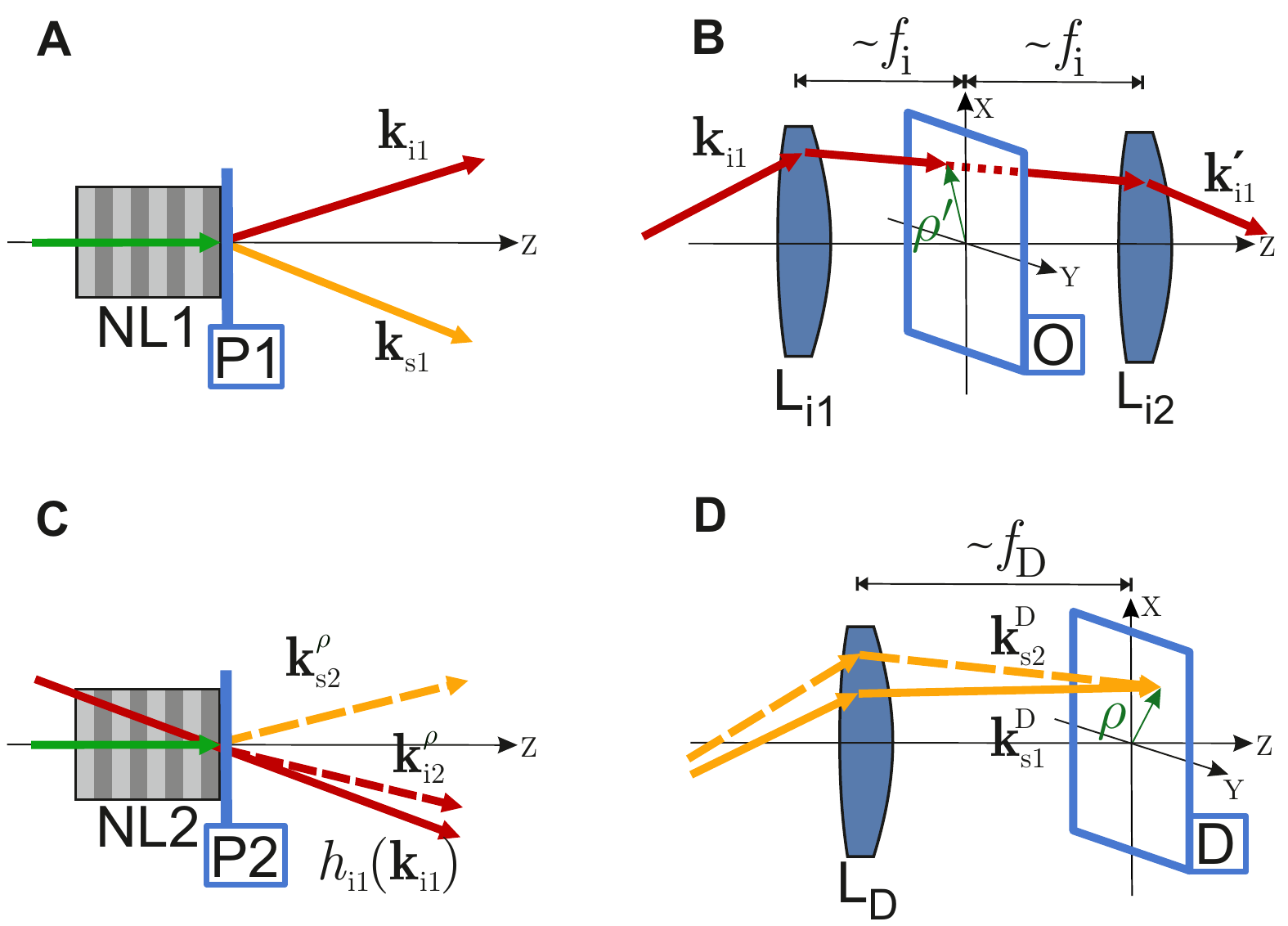}
    \caption{\label{fig:vector_relations}\textbf{Sketch illustrating the notation at different parts of the setup.} Panels (\textbf{A}) and (\textbf{C}) show the origin of the biphoton states. The relation of $\kia$ to a point $\rho^{\prime}$ in the object plane is illustrated in panel (\textbf{B}). The set of $\textbf{k}_{\mathrm{s}}$-vectors in (\textbf{A}) and (\textbf{C}) is chosen such that the positions $\rho$ match in the detector plane as shown in panel (\textbf{D}). The $\textbf{k}$-vectors are deliberately not parallel before the lenses to demonstrate the case of a misaligned system denoted by the distances ${\sim} f_\mathrm{i}$.}
\end{figure} 

Next we consider the propagation, which encompasses the change of momenta and positions, losses, and phase-shifts of the biphoton states. We model losses due to propagation and from the imaged object as beam splitters with transmittance $t$ and a phase shift $\phi$ such that they can be described by the transformation $\adagger \rightarrow t \mathrm{e}^{i\phi}\adagger + \sqrt{1-t^2}\adagger_{\mathrm{loss}}$. Here, $\adagger$ are the photon creation operators and $\adagger_{\mathrm{loss}}$ represents all losses along a beam path or from the object. In general, the transmittances and phase shifts depend on the photon momenta. We omit this dependence in the following equations for the sake of readability. The propagation of the states $\ket{\psi_{\mathrm{NL1/NL2}}}$ through the beam paths from a source plane to the detectors plane is modeled by the transformation of the photon momenta $\mathbf{k}_{m} \rightarrow \mathbf{k}_{m}^{\mathrm{D}} = h_{m}(\mathbf{k}_{m})$ for $m = \mathrm{i1,i2,s1,s2}$. Here $h_{m}$ is a transformation function modeling the changed photon momenta after the propagation from the source planes P1 and P2 to the detector plane D. The definition of the planes is shown in Fig.~2. 

In an ideal setup, a plane wave emerging from one of the crystals is focused onto a point on the detector \cite{Lahiri2015}. This point $\rho(\ks)$ depends on the momentum vector $\ks$ which defines the plane wave. This is illustrated in Fig.~2D. The same holds for the relation of momentum vector $\ki$ and a point on the object $\rho^{\prime}(\ki)$ (Fig.~2B). 
For a more realistic setup, these assumptions do not hold as the components are not perfect and further might be misaligned. 
In order to still have a momentum-to-point relation, we make the assumption that the spatial extent of the sources can be neglected.

With this, we can establish a relation between the photons from the two sources (Figs.~2A and C). Fig.~2 illustrates these relations for a Mach-Zehnder setup as well as the employed notation. A biphoton state emitted from NL1 with signal momentum $\ksa$ is detected at the point $\rho(\ksa)$ on the detector. For a biphoton state from NL2 to be detected at the same point the signal part needs to have the momentum $\ksb^{\rho}$ in the source plane NL2 which is the momentum that fulfills $\rho(h_{\mathrm{s2}}(\ksb^{\rho})) = \rho(\ksa^{\mathrm{D}})$. We can establish an analogous relation for the idler photons. With these relations from the propagation model we can formulate the detection model. The quantum state at the detector plane can be expressed as

\begin{align}
    \label{eq:detector_state}
    \ket{\psi_{\mathrm{D}}} =  
    \int \dr\ksa & \dr\kia\left[ g_{1} (\ksa, \kia)t_{\mathrm{s1}} t_{\mathrm{i}} t_{\mathrm{o}} \mathrm{e}^{i\Delta \phi_1} \adagger(\ksda)\adagger(\kida) \right.   \nonumber\\
    & +g_{1} (\ksa, \kia)t_{\mathrm{s1}} \sqrt{1- (t_{\mathrm{i}} t_{\mathrm{o}}) ^2} \mathrm{e}^{i\Delta \phi_2} \adagger(\ksda)\adagger_{\mathrm{i,loss}}  \nonumber\\
    & +g_{1} (\ksa, \kia)t_{\mathrm{i}} t_{\mathrm{o}} \sqrt{1- t_{\mathrm{s1}}^2} \mathrm{e}^{i\Delta \phi_3} \adagger_{\mathrm{s,loss}}\adagger(\kida) \nonumber\\
    & + \left.g_{2} (\mathbf{k}_{\mathrm{s2}}^{\rho}, \mathbf{k}_{\mathrm{i2}}^{\rho}) t_{\mathrm{s2}} \mathrm{e}^{i\Delta \phi_4} \adagger(\ksdb)\adagger(\kidb)\right] \ket{0}
\end{align}
where the $\Delta \phi_n$ denote the phase shifts from the propagation through the various optical paths which are provided in detail in Eqs.~10 to 13, $t_{\mathrm{o}}$ is the transmittance of the imaged object, $t_{\mathrm{s1/s2}}$ are the transmittances of the signal paths between source and detector planes and $t_{\mathrm{i}}$ is the transmittance of the idler path between the two source planes. The first three terms describe the biphoton states emitted from crystal NL1. Starting with the case where both photons are transmitted through the system and then the cases where only signal or only idler photons are transmitted. The fourth term describes biphotons from the second crystal. We omitted further terms which do not contribute to the measured signal such as idler losses after NL2.

From the state at the detector we obtain the count rate for a point on the detector plane as
\begin{align}
\label{eq:countrate}
    \Gamma^\pm (\rho(\ksda))& =   
    \bra{\psi_{\mathrm{D}}} \left[\adagger(\ksda) +\adagger(\ksdb)\right] \left[\hat{a}(\ksda) + \hat{a}(\ksdb)\right]\ket{\psi_{\mathrm{D}}} \\
           & = \int \dr\kia \left[ g_1 (\ksa,\kia)^2 t_{\mathrm{s1}}^2 + g_{2} (\mathbf{k}_{\mathrm{s2}}^{\rho}, \mathbf{k}_{\mathrm{i2}}^{\rho})^2 t_{\mathrm{s2}}^2 \right. \nonumber\\
            & \left. \quad\quad\pm 2g_1 (\ksa,\kia) g_{2} (\mathbf{k}_{\mathrm{s2}}^{\rho}, \mathbf{k}_{\mathrm{i2}}^{\rho}) t_{\mathrm{s1}} t_{\mathrm{s2}} t_{\mathrm{i}} t_{\mathrm{o}} 
             \cos(\Delta \phi) \right]
\end{align}
where $\Delta \phi =\Delta \phi_1 - \Delta \phi_4$ is the phase difference of biphoton states created in NL1 and NL2. In general, the phase terms depend on the photon momenta and therefore the cosine term cannot be extracted from the integral. The different signs of the third term stem from the $\pi$ phase shift the beam splitter introduces to one of its output ports.
To obtain the count rate for a detector pixel we integrate $\Gamma^{\pm}$ over the area covered by this pixel and multiply with the detector efficiency.

From the measurements at the two outputs (see Fig.~1), we can calculate the pointwise visibility which gives an approximation of the optical properties of the object in the idler path:
\begin{align}
    \label{eq:vis1}
    V(\rho(\ksda)) =& 
    [\Gamma^{+}(\rho(\ksda))-\Gamma^-(\rho(\ksda))]/[\Gamma^{+}(\rho(\ksda))+\Gamma^-(\rho(\ksda))] \\
    \label{eq:vis_explicit}
    =&\frac{\int \dr\ki 2g_1(\ksa,\kia) g_{2}(\mathbf{k}_{\mathrm{s2}}^{\rho},\mathbf{k}_{\mathrm{i2}}^{\rho}) t_{\mathrm{s1}} t_{\mathrm{s2}} t_{\mathrm{i}} t_{\mathrm{o}}
    \cos(\Delta\phi)}
    {\int \dr\ki [ g_1 (\ksa,\kia)^2 t_{\mathrm{s1}}^2 + g_{2} (\mathbf{k}_{\mathrm{s2}}^{\rho}, \mathbf{k}_{\mathrm{i2}}^{\rho})^2 t_{\mathrm{s2}}^2]}
\end{align}
where we have assumed identical imaging systems for the two measurements. In this general case, the relation of this quantity and the object properties is not straightforward. For an idealized setup with perfect momentum correlation, the idler momentum integrals vanish which yields $V \propto t_{\mathrm{o}}
 \cos(\phi_{\mathrm{o}})$ \cite{Lahiri2015}. 
We can make some approximations to simplify Eq.~\ref{eq:vis_explicit} while still retaining important properties of the system. For this we assume the crystals to be identical ($g = g_1 = g_2$), constant and equal losses in the optical paths ($t_{\mathrm{s}}= t_{\mathrm{s1}} = t_{\mathrm{s2}}$), vanishing phases ($\Delta \phi = \phi_{\mathrm{o}}$), and an optical system that ensures $\ksa = \ksb^{\rho}$ and $\kia = \kib^{\rho}$. This allows us to reduce Eq.~\ref{eq:vis_explicit} to: 
\begin{equation}
    \label{eq:visibility_reduced}
     V(\rho(\ksda)) = \frac{\int \dr \ki g(\ksa,\kia)^2 t_{\mathrm{s}}^2 t_{\mathrm{i}} t_{\mathrm{o}} \cos(\phi_{\mathrm{o}})}{\int \dr \ki g(\ksa,\kia)^2 t_{\mathrm{s}}^2}
\end{equation}
where only the transition probabilities and object properties $t_{\mathrm{o}}$ and $\phi_{\mathrm{o}}$ depend on the photon momenta. 

We further simplify and consider biphoton sources where the normalized transition probabilities $\hat{g} = g(\ks,\ki)^2/\int \dr \ki g(\ks,\ki)^2$ depend only on the difference in $\ks$ and $\ki$, i.e. $\hat{g} = \hat{g}(\ks-\ki)$. In the case of a purely absorptive object ($\phi_{\mathrm{o}} = 0$) this reduces Eq.~\ref{eq:visibility_reduced} to a convolution of the objects transmittance with the $\hat{g}$ kernel:
\begin{equation}
    \label{eq:visibility_convolution}
     V(\rho(\ks^{\mathrm{D}})) = t_{\mathrm{i}}\int \dr \ki \hat{g}(\ks-\ki) t_{\mathrm{o}}(\rho^{\prime}(\ki))
\end{equation}
where the dependence of the object transmittance is now explicitly given. Due to the constant convolution kernel this approach yields a computationally fast method for calculating the measurement for any object. From this, properties of the imaging setup such as resolution limit and maximum visibility can be inferred. 
The employed assumption can readily be applied if the region of interest is the center of the spots in the detector plane or when the higher relative noise at lower count rates can be neglected. 
The assumption of vanishing phases can be relaxed giving a more general case where the convolution kernel becomes shift variant as it also contains the cosine term. Including the signal $\phi_{\mathrm{s}}$ and idler phases $\phi_{\mathrm{i}}$ allows for the emulation of misalignments, such as transversally shifted lenses in the two optical arms. An improvement in computational efficiency is retained as long as $\hat{g}$ needs to be evaluated only sparsely. Both the quasi-Monte Carlo simulation and the convolution approximation can be extended to a broader spectral range by integrating over $\ksa$ as well. 

The insight, that the measured image can be seen as a convolution of the object combined with a way to calculate the convolution kernel, paves the way for using established deconvolution and image reconstruction techniques. Using these techniques in the post-processing of the detected images allows to improve their quality without any improvements in the detection system. Applied correctly, deconvolution techniques provide a more accurate approximation of the inferred object properties. To demonstrate this, we use the Richardson-Lucy algorithm \cite{Richardson1972,Lucy1974} to deconvolve the simulated measurement results with the convolution kernel $\hat{g}$. 

\subsection*{Implementation}
To demonstrate the capabilities of our simulation methods, we model a far-field Mach-Zehnder interferometer setup presented in \cite{Lemos2014,Fuenzalida2022}. The setup's sources create biphoton states with central wavelengths $\lambda_\mathrm{s} = 810\,$nm and $\lambda_\mathrm{i} = 1550\,$nm. A detailed account of the model of the setup is given in the Materials and Methods section.

To describe the biphoton states produced by the two spontaneous parametric downconversion (SPDC) sources NL1 and NL2 we use a highly accurate and general model for predicting absolute photon rates \cite{Riexinger2021}. This sophisticated model takes many of the material properties into account, such as the wavelength dependence of refractive indices and nonlinear susceptibility yielding an accurate and complex SPDC model that needs to be evaluated numerically. Note that the approximations that have been made for the derivation of the SPDC model do not limit the generality of the presented method, as it can be substituted with a different source model. 

The magnification of the setup depends on the wavelengths of signal and idler \cite{Lemos2014,Lahiri2015}, it can be approximated by
\begin{equation}
\label{eq:magnification}
    M = \frac{f_\mathrm{D} \lambda_\mathrm{s}}{f_\mathrm{i} \lambda_\mathrm{i}}
\end{equation}
In general, the magnification is different along the crystal axis due to the birefringence of many nonlinear crystals. For the setup presented in this paper, these differences can be neglected.

The transformation functions $h_{m}$ are modeled using geometrical optics. The momentum $\mathbf{k}$ and position $\rho$ in the plane of interest is evaluated by propagating a ray defined by the photon momentum to the corresponding plane. This is used to determine the position on the object and detector, as well as the photon momenta in the plane of the second source $\mathbf{k}_{\mathrm{s2}}^{\rho}$. The phase terms are determined from the optical path lengths $l_m$ as $\phi_{m} = \omega_{m} l_{m}/c$. For the aligned interferometer we assume that only the phase shift from the object is relevant, i.e. $\Delta \phi = \phi_{\mathrm{o}}$. We further assume that no losses occur in the optical paths ($t_{\mathrm{s}} = t_{\mathrm{i}} = 1$), as these are either constant factors which reduce the visibility. Or they are complex functions which would diminish the clarity of the demonstration of our simulation method. 
For a full simulation of the QIUP setup we evaluate Eq.~\ref{eq:countrate} over the detector area using a quasi-Monte Carlo integration scheme. 
For the convolution variant, we evaluate the normalized transition probabilities $\hat{g}(\ks-\ki)$ for the collinear $\ks$ defined by the optical axis of the system and the central signal wavelength of the setup $\lambda_\mathrm{s}$. This is used to calculate the convolution of the object given in Eq.~\ref{eq:visibility_convolution}. 
The computational expense of evaluating $\hat{g}$ is equivalent to the evaluation of one pixel on the detector, reducing the theoretical computation time by six orders of magnitude for this setup. 

\begin{figure}[ht]
    \centering
    \includegraphics[width=0.5\textwidth]{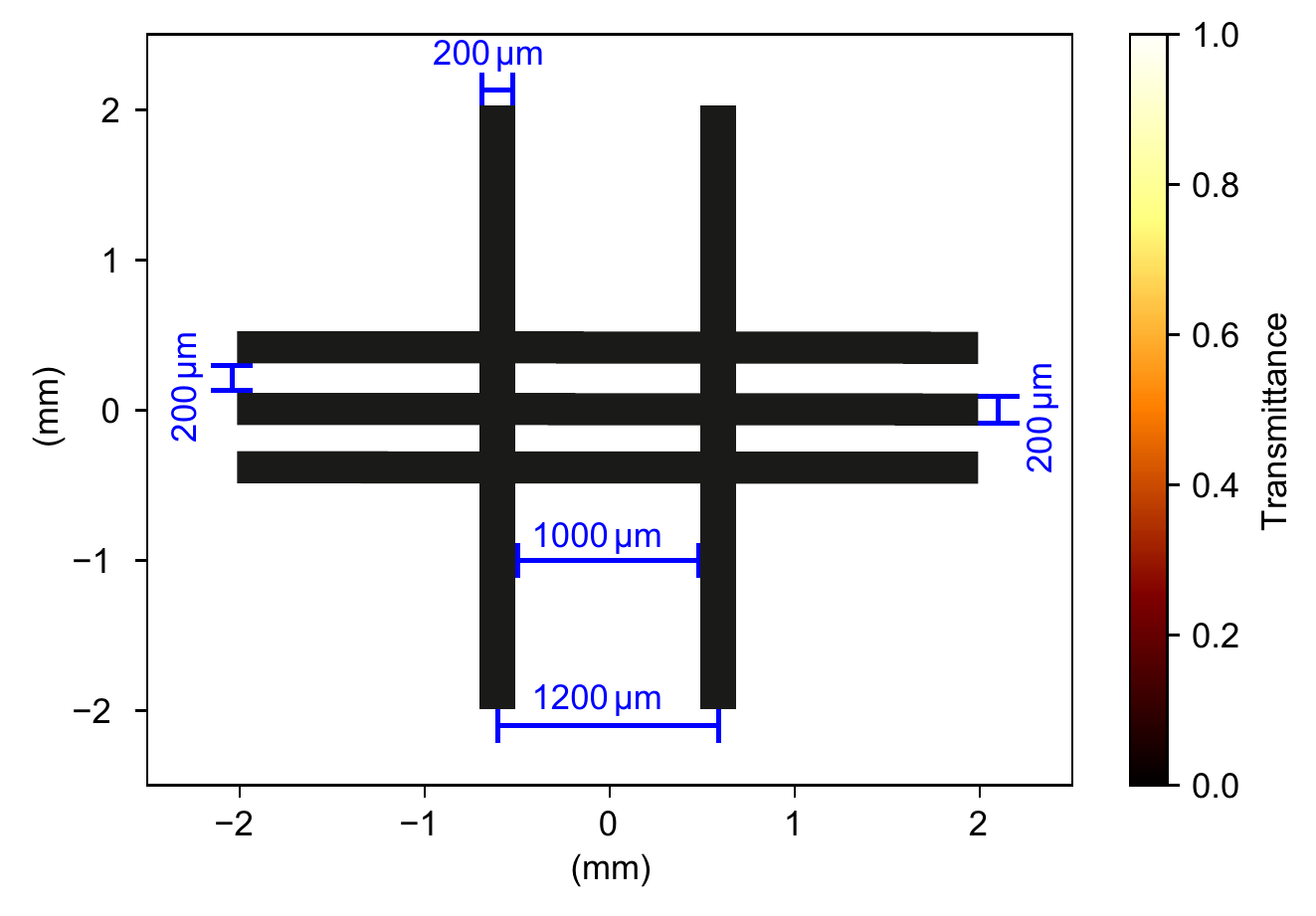}
    \caption{\label{fig:object}\textbf{Transmission object used for the simulations.} The object measures 4$\times4$\,mm$^2$ and consists of horizontal and vertical opaque bars at different distances.}
\end{figure} 

\subsection*{Numerical Results}

\begin{figure*}
    \centering
    
    \includegraphics[width=\textwidth]{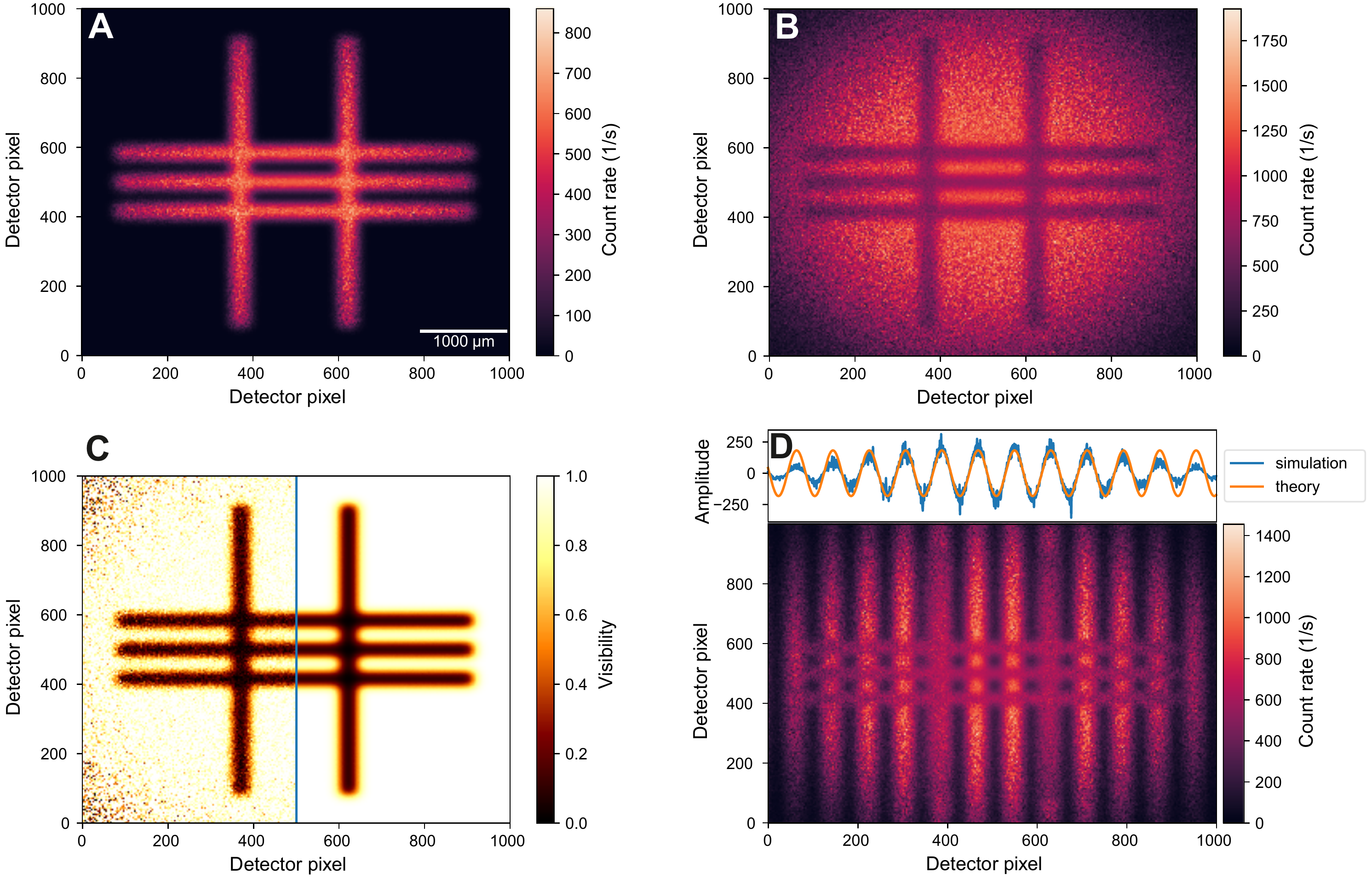}
    \caption{\textbf{Simulation results.} Simulated measurements of the transmission object shown in Fig.~3. (\textbf{A}) and (\textbf{B}): Detector images for the destructive ($\Gamma^{-})$ and constructive ($\Gamma^{+}$) configuration. The noise stems from the finite sampling rate and is perfectly correlated between the two images. (\textbf{C}): The resulting measured transmittance. For the left half Gaussian noise was added to the detector images, the right half is free from artificial noise. (\textbf{D}): Constructive detector image for the misaligned setup. The lens misalignment leads to vertical stripes in the detector plane. The top panel shows simulated and theoretical modulation from the phase shift at $\mathrm{y}=50$.}
    \label{fig:imagingExample}
\end{figure*}

We simulate the detector images with the absorption object shown in Fig.~3. It measures \mbox{$4\times4$\,mm$^2$}, has three horizontal, $200$\,µm wide bars with the same distance between them, and two vertical bars of the same width but separated by $1000$\,µm. The pump beam waist used for this simulation is $300$\,µm. The results of the simulated detector images and measurement of the transmittance of a test object are shown in Fig.~4. The results for the detector count rates $\Gamma^-$ and $\Gamma^+$ are shown in Figs.~4A and 4B, respectively. The two simulated images show the expected behavior, where the sum of both images gives the intensity profile of the signal radiation without distinctive features, whereas the difference shows an image of the object. These figures also show the spot size of the signal radiation. 
The noise visible in the simulated detector images is a numerical artifact which is caused by the finite number of points we evaluated. The noise is perfectly correlated between the two images such that the resulting pointwise visibility shows no artifacts from this noise. 
Fig.~4C shows the measured pointwise visibility of the object which is defined by Eq.~\ref{eq:vis1}. To demonstrate more realistic conditions, we add Gaussian noise with a standard deviation of $50$ counts to the left half of the detector images for the calculation of this image. We clip values below $0$ and above $1$.
The right half is free from artificial noise to illustrate the influence of this added noise. The resulting measured visibility shows a smeared out image of the object. The visibility in the noise-free half of the image ranges from $0$ to $1$ as we neglected misalignment and losses from the optical paths and components which would reduce it. The visibility does not decrease towards the outer parts of the features. In the noisy half of the image, the visibility ranges from $0.1$ to $1$ and the noise is more prominent in the corners where the count rate is lower. 
The measured distance of the centers of the vertical bars is $1255$\,µm. The distance on the object is $1200$\,µm (see Fig.~3) which gives a magnification value of $1.046$, in agreement with the theoretical prediction from Eq.~\ref{eq:magnification} ($M = 1.045$). The accuracy of the measured magnification is limited by the detector pixel size.

% Misaligned Setup
In real experimental setups, noise is not the only issue. Even with included noise, they will always fall short of this ideal setup. As components have imperfections and cannot be aligned perfectly, our model needs to be able to reflect these effects. Due to the flexible propagation model, this is possible. To demonstrate our method's capabilities for misaligned systems, we shift the idler lens L$_{\mathrm{i1}}$ in front of the object by $d_{\mathrm{t}} = 0.3$\,mm transversally respective to the optical axis as shown in Fig.~1. This introduces a phase shift which depends on the emission angle of the idler photons. Fig.~4D shows the introduced vertical stripes in the simulated detector images as one would expect from misaligning an interferometer. The width of the stripes depends on the magnification and the misalignment. An analytical expression for the phase shift in the detector plane is derived in Eq. 17. The analytical and simulated results are shown in the top panel of Fig.~4D. Our simulated stripe pattern matches the theoretical predictions closely.
The pointwise visibility in this setup is decreased to the range from $0$ to $0.64$ due to the phase variation in the idler states. 

We further calculated the resolution limit for both setups numerically (see supplementary materials) to compare with the measured results of Fuenzalida \textit{et al.} \cite{Fuenzalida2022}. Our results show the same characteristic curve, but are consistently smaller than the measured results for both experimental setups. Our simulated resolution limits are $\sim$\,$17$\% smaller.
This is to be expected when comparing an ideal with a real setup. Reasons for the differences are the common difficulties in real experiments: imperfect overlap of the beam paths, losses, spherical aberrations, noise and others. Our results present a theoretical limit to the achievable resolution for these setups and show that these experimental realizations might still be improved. A detailed analysis of the experimental setups can be performed using our simulation method. This allows the identification of sensitive parts and limiting factors, which can be used to improve the setups. 

\begin{figure}
    \centering
    \includegraphics[width=0.5\textwidth]{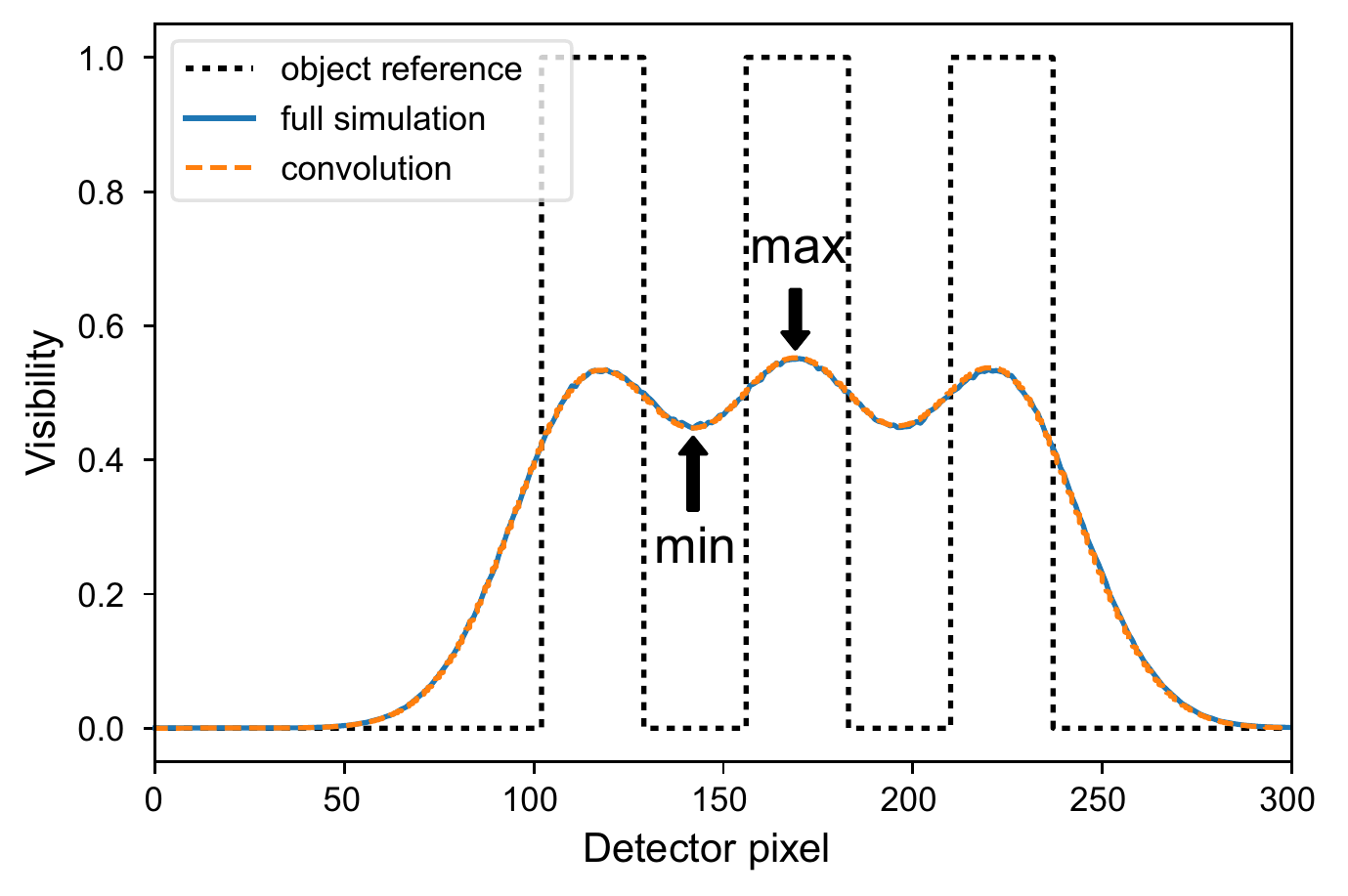}
    \caption{\textbf{Comparison of simulation methods.} The figure shows the measured transmittance from the full numerical simulation (blue) and the convolution method (orange). The perfect image of the object is given as a reference (black).}
    \label{fig:sim_conv}
\end{figure}

% Convolution
With these examples we have demonstrated the capability and versatility of the presented simulation method. Under certain conditions we can formulate the pointwise visibility as a convolution, which leads to a huge speedup of the simulation times while still giving accurate results. To demonstrate this, we change the beam waist to $200$\,µm and use an absorption object that consists of three vertical slits with transmittance $t=1$. The width and distance between the slits is $128$\,µm which is the resolution limit for the setup in this configuration. A horizontal cut at the center ($y=500$) of the detector is shown in Fig.~5. The figure shows the results for the full simulation and the convolution method. The convolution method yields an accurate approximation of the fully simulated image. The maximum difference in visibility over the shown area is $0.007$, showing that this method provides a fast and accurate approximation for the full simulation results. An extension of this method to phase objects and misaligned setups and the corresponding simulations are shown in the supplementary materials. A downside of this method is that the limited spot size is not considered. For system characteristics such as visibility and resolution limit the convolution method provides a fast calculation method that enables the exploration of the influence of different parameters and the optimization of setups.

\begin{figure}
    \centering
    \includegraphics[width=0.5\textwidth]{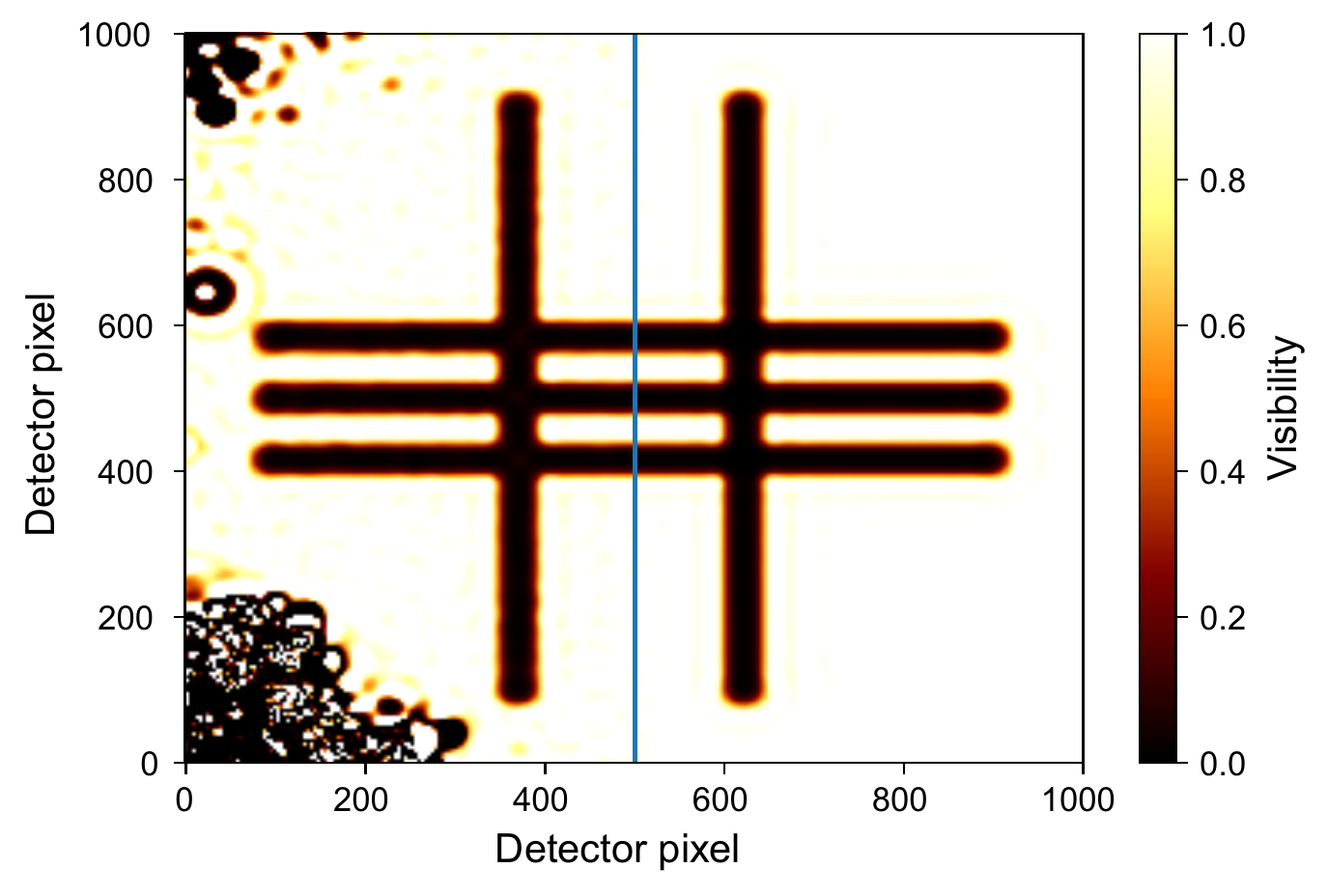}
    \caption{\textbf{Deconvolution results.} Shown is the deconvolution of the simulated visibility of Fig.~4C using the normalized transition probabilities $\hat{g}$ of the system. The image becomes sharper and the noise is notably reduced. The root mean square error to the real object is improved by $37$\% and $35$\% in the noisy and noise-free half.}
    \label{fig:deconvolved}
\end{figure}

% Deconvolution
The convolution kernel used for the convolution variant can also be used to improve the information we obtain from the two measurements. For this we apply the Richardson-Lucy deconvolution with the collinear convolution kernel $\hat{g}$ to the simulated visibility shown in Fig.~4C. The result after 50 iterations, shown in Fig.~6, is a much sharper image with narrower edges and an increased contrast, especially in the center. Outside of the beam spot the large relative noise leads to large artifacts. We therefore limit our analysis to a box defined by $x\in[100,900]$ and $y\in[300,700]$ which cuts off the parts where noise is dominant. In this box, the deconvolution algorithm introduces only minor artifacts. With a maximum amplitude of $0.05$, they are barely visible in the figure. In this area, the noise leads to minor distortions. The addition of noise does negligibly increase the amplitude of the artifacts compared to the noise-free half.
To compare the deconvolution to the regular simulation results, we calculate the root mean square error (RMSE) to the real object for both methods. The deconvolution method improves the RMSE in the noisy and noise-free half by $37$\% and $35$\%, respectively. The range for the pointwise visibility in the center of the image is improved from [$0.1,\,0.8$] to [$0,\,1$], which is the maximum achievable contrast. In the noise-free half, the maximum improvement for a single pixel is $0.44$, the mean improvement on this side is $0.1$.
To quantify the improvement in the sharpness, we have calculated the resolution limits including the deconvolution step. Compared to the simulations without deconvolution, we found a consistent improvement of 13\% across a large range of beam waists.

\section*{Discussion}

We have presented and demonstrated a general method for simulating quantum imaging with undetected photons. For the first time, these simulations enable the characterization of QIUP systems in the design phase. 
We presented a method for full fledged simulations of QIUP systems which allows analyzing a large range of setups in full detail, including the influence of many misalignments. The convolution approximation provides a trade-off between computation time and accuracy and also a fast way to evaluate characteristics of a setup with slightly limited options for misalignments. The numerical simulations enable the identification of limiting factors of QIUP systems. The fast computation times even allow for an interactive exploration of the design space and optimization of the imaging characteristics -- an invaluable tool for improving applications based on QIUP. Both methods yield similar results for perfect imaging systems. A comparison with analytical and experimental results shows that the simulation results show realistic benchmarks of experimental setups. 

We further demonstrated the improvement of imaging characteristics by deconvolution techniques. In our simulation experiment, we showed that the quality of an image can be improved greatly by applying this technique.

Our results demonstrate the principle of the methods. Using more accurate optical models to reproduce the images from experimental setups is subject of further research. The convolution technique can also be improved. By incorporating multiple convolution kernels, it is possible to get a better approximation for the variation in different detector regions, allowing for an estimation of the absolute count rate as well. 
Further work is also possible in applying the deconvolutional technique to real experimental setups and data. This requires an accurate model of the setup and its imperfections. Using a deconvolution algorithm for shift-variant kernels allows adding the phase information to the deconvolution method which can improve the image further, since in real setups phase errors always occur.

The presented approach can also be applied to a wide range of other experimental setups. We have presented an approach specifically for collinear spectroscopy experiments before \cite{Kutas2020}. The application to a Michelson geometry or a near-field configuration can be realized by adapting the transfer functions \cite{Gilaberte2021} or by using position-space operators instead of momentum-space operators \cite{Viswanathan:21}, respectively.
The simulation method we present can be further extended to include the spatial component of the biphoton sources. However, this requires a different source model and is beyond the scope of this proof-of-concept paper.
At the current stage, the presented model and methods already provide the necessary basis for the simulation of experiments and applications based on quantum imaging with undetected photons and enables obtaining improved images. 

\section*{Materials and Methods}
\subsection*{Phase terms}\label{sec:phase_terms}
The phase terms $\Delta\phi_n$ in Eq.~\ref{eq:detector_state} are composed of the phase shifts from the propagation of the biphoton states through the optical system. These can be separated into individual components $\phi_m$. Where the indices $m=\mathrm{p1,p2}$ denote the propagation of the pump laser to the source planes, $m=\mathrm{s1,i1,s2,i2}$ denote the phase shift of propagation from source to detector plane and ${\mathrm{o}}$ denotes the phase shift introduced by the object. The phase terms are then defined as
\begin{align}
    \Delta\phi_1 &= \phi_{\mathrm{p1}} + \phi_{\mathrm{s1}} + \phi_{\mathrm{i1}} + \phi_{\mathrm{o}} \\
    \Delta\phi_2 &= \phi_{\mathrm{s1}} + \xi_{\mathrm{i1}} + \xi_{\mathrm{o}} \\
    \Delta\phi_3 &= \xi_{\mathrm{s1}} + \phi_{\mathrm{i1}} + \phi_{\mathrm{o}} \\
    \Delta\phi_4 &= \phi_{\mathrm{p2}} + \phi_{\mathrm{s2}} + \phi_{\mathrm{i2}}
\end{align}
where $\xi_m$ stand for phase shifts which depend on the position where the photon gets absorbed. In our model the precise form of $\xi_m$ is not further specified as these terms do not influence the detected signal.
Using the form of these terms we find the phase shift modulating the image
\begin{equation}
    \Delta\phi = \phi_{\mathrm{o}} + \phi_{\mathrm{p1}} + \phi_{\mathrm{s1}} + \phi_{\mathrm{i1}} - \phi_{\mathrm{p2}} - \phi_{\mathrm{s2}} - \phi_{\mathrm{i2}}
\end{equation}
which shows that we observe constructive interference when $\Delta\phi$ is an even multiple of $\pi$ and destructive interference when it is an odd multiple.

\subsection*{Experiment models} The setup models we evaluate are based on the experimental setups presented by Fuenzalida \textit{et al.} \cite{Fuenzalida2022}. These setups have a Mach-Zehnder geometry and share the same optical components. The crystals used are periodically poled potassium titanyl phosphate (PPKTP) crystals of the dimensions $2\times2\times1$\,mm$^3$ (L$\times$W$\times$H) with a type-0 phase-matching scheme. The setups differ in the crystal poling periods $\Lambda_{1/2}$ and pump lasers which results in different signal and idler wavelengths. The first setup, which is used for the results in the main text, contains PPKTP crystals with $\Lambda_1 = 9.675$\,µm which are pumped by a \mbox{$532$\,nm laser}. The second setup contains crystals with $\Lambda_2 = 5.33$\,µm which are pumped by a \mbox{405\,nm laser}. The first setup’s sources create biphoton states with central wavelengths $\lambda_\mathrm{s} = 810$\,nm and $\lambda_\mathrm{i} = 1550$\,nm. The second setup creates states with central wavelengths $\lambda_\mathrm{s} = 842$\,nm and $\lambda_\mathrm{i} = 780\,$nm.
The temperatures of the crystals are $85^\circ$C in the first setup and $75^\circ$C in the second setup. Both lasers are modeled with a power of $50$\,mW. We choose a detector pixel size of $5\times5$\,µm$^2$.
All lenses are modeled as ideal lenses defined only by their focal lengths. The focal lengths of the lenses are $f_{\mathrm{i}} =75$\,mm for L$_{\mathrm{i1}}$, L$_{\mathrm{i2}}$, L$_{\mathrm{s1}}$, and L$_{\mathrm{s1}}$. The lens L$_{\mathrm{D}}$ in front of the detector has a focal length of $f_{\mathrm{D}} = 150$\,mm. 
The theoretical value for the magnification is $M = 2.159$ for setup 2. As the focal lengths are the same in both setups, the difference stem from the ratio of wavelengths. The measured distance of the centers of the vertical bars is $2590$\,µm, which gives a magnification of $2.158$ in good agreement with the theoretical value.

\subsection*{Phase shift in misaligned setup}
The setup is misaligned by a transversal shift of lens L$_{\mathrm{i1}}$. To obtain an analytical expression for the resulting phase shift in the detector plane we calculate the path-length difference with ray optics. We start with the phase shift from a transversal displacement of a perfect lens in the parabolical approximation. The phase shift from a lens aligned on the optical axis is given by
\begin{equation}
    \phi_{\mathrm{L}}(x_{\mathrm{L}}) = \frac{\pi x_{\mathrm{L}}^2}{f \lambda}
\end{equation}
where $x_{\mathrm{L}}$ is the position where the ray passes through the lens, $f$ is the focal length of the lens and $\lambda$ is the wavelength of the light. Assuming that the imaging system is aligned before the lens displacement, only the additional phase introduced by the lens shift is relevant. This is given by
\begin{align}
    \Delta \phi_{\mathrm{L}} (x_{\mathrm{L}}) &= \phi_{\mathrm{L}} (x_{\mathrm{L}} + {\mathrm{t}}) - \phi_{\mathrm{L}}(x_{\mathrm{L}}) \\
    &= \frac{\pi {\mathrm{t}}^2+2\pi x_{\mathrm{L}}{\mathrm{t}} }{f \lambda}
     \label{eq:lens_phase_shift}
\end{align}
where ${\mathrm{t}}$ is the transverse lens shift. To obtain the phase shift in dependence of the position in the detector plane, we use the relation $x_{\mathrm{D}} = M x_{\mathrm{L}}$ which is valid for the two lenses L$_{\mathrm{i1}}$ and L$_{\mathrm{i2}}$ in the idler path.

% Your references go at the end of the main text, and before the
% figures. For this document we've used BibTeX, the .bib file
% scibib.bib, and the .bst file Science.bst. The package scicite.sty
% was included to format the reference numbers according to *Science*
% style.

%BibTeX users: After compilation, comment out the following two lines and paste in
% the generated .bbl file. 

\bibliography{scifile}

\bibliographystyle{Science}

\section*{Acknowledgments}
\paragraph*{Funding:} This project was funded by the Fraunhofer-Gesellschaft within the Fraunhofer Lighthouse Project Quantum Methods for Advanced Imaging Solutions (QUILT). 
\paragraph*{Author contributions:} G.v.F. and M.B. initiated this research. F.R. performed the theoretical and numerical analysis. All authors discussed the results and contributed to the writing of the manuscript. 
\paragraph*{Competing interests:} The authors declare that they have no competing interests. 
\paragraph*{Data and materials availability:} All data needed to evaluate the conclusions in the paper are present in the paper. Additional data related to this paper may be requested from the authors.

\newpage
\section*{Supplementary materials}
\setcounter{figure}{0}
\renewcommand{\thefigure}{S\arabic{figure}}

\subsection*{Resolution limit} 
We use the definition of the resolution limit given in ref.~\cite{Fuenzalida2022}. This definition uses an object with three slits of width $d$, separated by the same length. We define $R$ as the ratio of visibility values at the local minima between slits and the maximum at the center (see Fig.~5) and the minimum resolvable distance as the slit width $d_{\mathrm{limit}}$ for which $R$ is equal to $0.81$. 
For the numerical calculation of the resolution limit, we first define a routine to calculate the ratio $R$. For this, we evaluate the collinear convolution kernel of a given setup. This is used to calculate the pointwise visibility of slits with width $d$ using the convolution variant of our method. From the resulting image we calculate the visibility ratio $R$. The minimum resolvable distance $d_{\mathrm{limit}}$ is then computed by an optimization algorithm. 

\begin{figure}[h]
    \centering
    \includegraphics[width=0.5\textwidth]{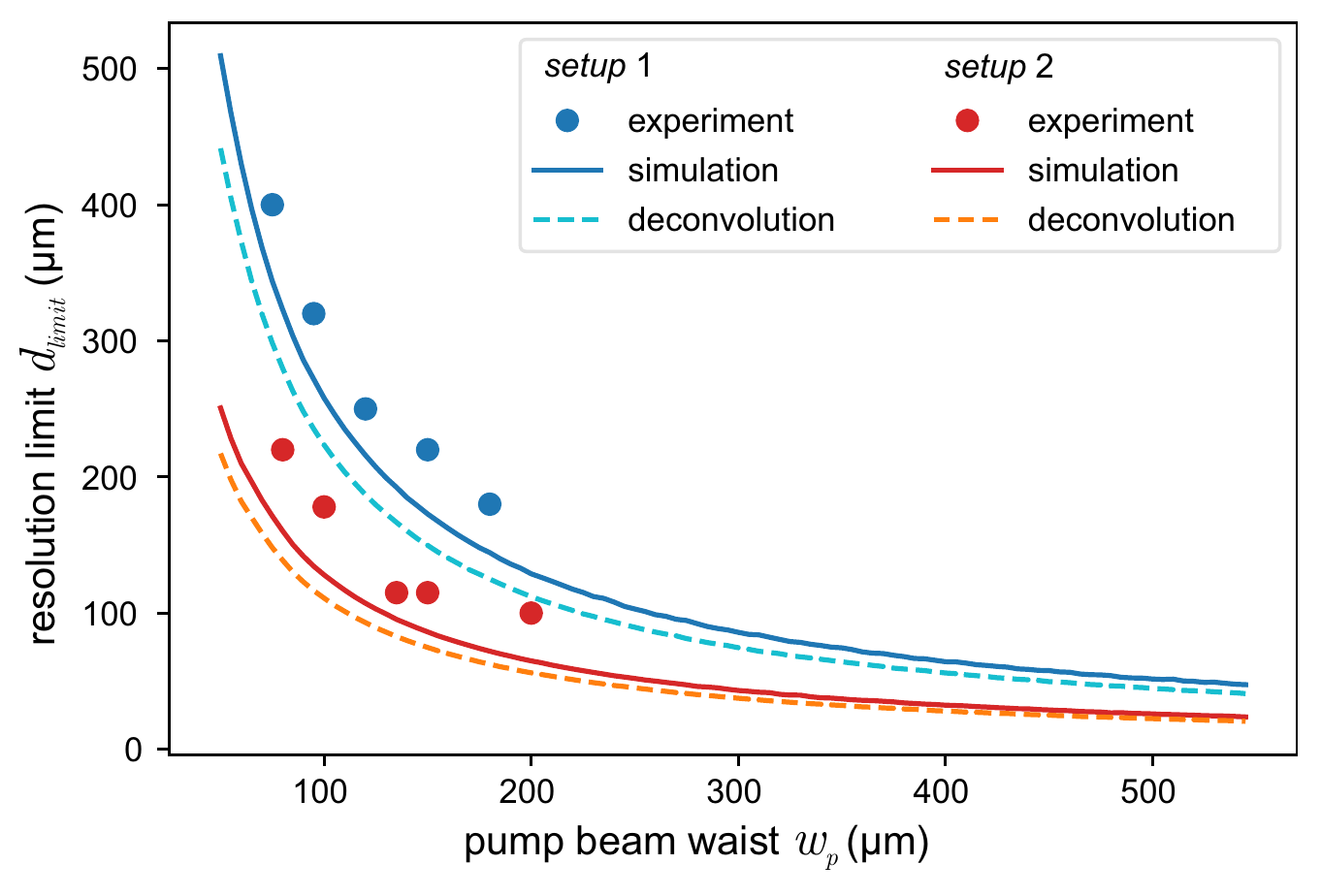}
    \caption{\textbf{Resolution limits.} Experimental and theoretical resolution limits for both setup 1 and setup 2. The experimentally measured values \cite{Fuenzalida2022} are larger than the simulation of the ideal setups. The deconvolution improves the resolution limit by $13$\% for both setups.}
    \label{fig:resolution_limt}
\end{figure}

This process is repeated with various pump beam waists $w_{\mathrm{p}}$ for both setup 1 and 2. In a similar fashion, we evaluate the resolution limit when a deconvolution is applied to the image before calculating $R$. The resulting resolution limits are shown in Fig.~S1. The experimental results are limited not only by measurement accuracy, but also because the used test target only provides discrete slit widths. The simulated results provide continuously calculated values.

\subsection*{Extension of convolution approximation}
The cosine term in Eq.~\ref{eq:visibility_reduced} vanishes under the assumption of an ideal setup and an absorption object. The convolution method can still be applied when these assumptions are relaxed. For this, we model the phase shift from the object and misalignments affecting the phase. An example of such a misalignment is defined by Eq.~\ref{eq:lens_phase_shift}. The phase shifts in this model depend on the locations and momenta at certain stages in the system. The locations can be determined from the momenta in the source planes. The resulting term for the visibility is similar to Eq.~\ref{eq:visibility_reduced}, but with the phase term $\phi_{\mathrm{o}}$ replaced by
\begin{equation}
    \Delta \phi = \sum_m \phi_m(\mathbf{k}_m, x_m)
\end{equation}
where $\phi_m$ are the phase shifts introduced by the object and the optical system, $\mathbf{k}_m$ are the momenta and $x_m$ are the transverse positions at the corresponding locations.
The resulting equation can be interpreted as a convolution with a shift-variant kernel, i.e. for each point $\rho(\ks)$ in the detector plane a different kernel has to be used. This slightly increases the complexity of the computations, but the resulting images can still be calculated within seconds. 
\begin{figure}[htb]
    \centering
    \includegraphics[width=\textwidth]{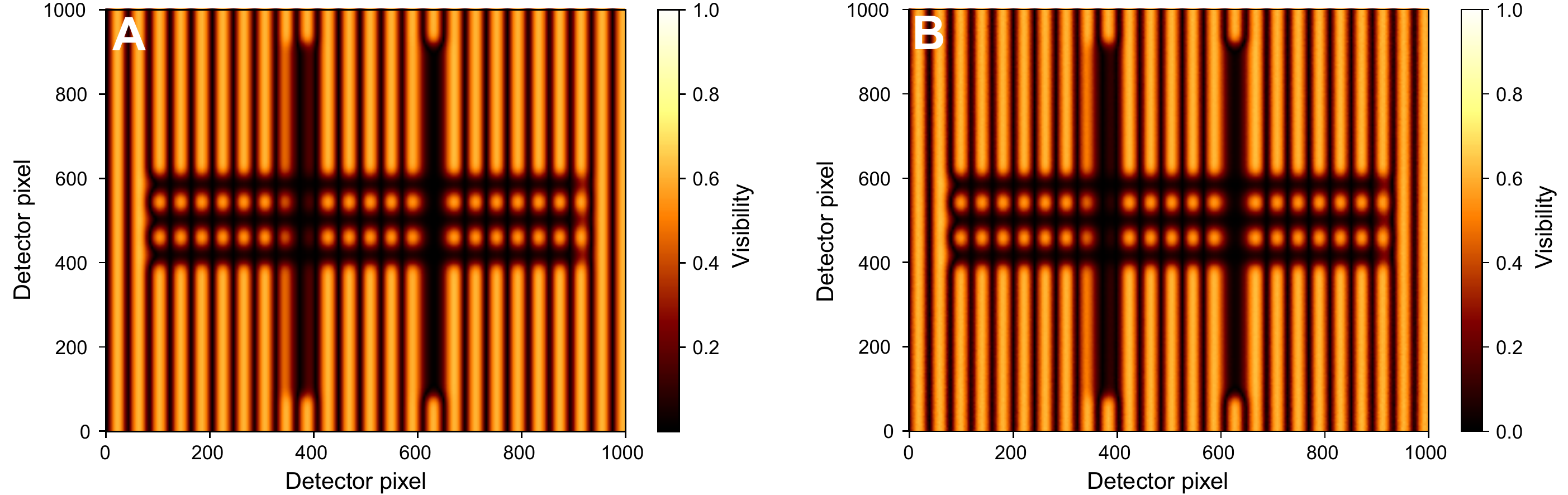}
    \caption{\textbf{Comparison of simulation methods for misaligned system.} The simulated visibility for the convolution approximation (\textbf{A}) and the quasi-Monte Carlo simulation (\textbf{B}).The transversely shifted lens leads to the vertical stripes in the measured visibility. The maximum visibility is reduced to $0.63$ due to the misalignment.}
    \label{fig:convolution_phase}
\end{figure}

We apply this method to the misaligned setup used for Fig.~4D. The resulting simulation of the measured visibility is shown in Fig.~S2. The image closely resembles the result from the quasi-Monte Carlo simulation. The difference of the two simulations is characterized by a RMSE of $0.013$ and a maximum error of $0.07$ over all pixels. The differences are larger than in the case without phases. This is explained by the more dynamic image caused by the stripes. Due to this, numerical inaccuracies such as in the magnification lead to larger errors.

\end{document}